\documentclass[aps,prc]{revtex4}

 \usepackage{epsfig}
 \usepackage{graphicx}
 \usepackage{dcolumn}
 \usepackage{amsfonts}
 \usepackage{amssymb}
 \usepackage{amsmath}
 \usepackage{mathrsfs}
 \usepackage{color}
 \newcommand{\blue}{\color{blue}}

 \newcommand{\black}{\color{black}}
 \newcommand{\beq}{\begin{equation}}
 \newcommand{\eeq}{\end{equation}}
 \newcommand{\beqa}{\begin{eqnarray}}
 \newcommand{\eeqa}{\end{eqnarray}}

 \newcommand{\bal}{\begin{align}}

\newcommand{\+}{\oplus}

\newcommand{\sdsum}{\mbox{$\,+\hspace{-0.40cm} \subset\;$}}  

\DeclareRobustCommand{\scrH}{\mathcal{H}} 

     {\begin{list}{}{%
     \settowidth{\labelwidth}{\textsf{#1:}}%
     \setlength{\leftmargin}{\labelwidth \labelsep -7pt}}}%
     {\end{list}}

 \def\Hb{\mathbb{H}}
 \def\Rb{\mathfrak{R}}
 \def\Cb{\mathbb{C}}
 \def\Bb{\mathbb{B}}
\def\la{\left\langle}
\def\ra{\right\rangle}

\begin{document}

 \title{Applications of the Capelli identities in physics and representation theory.\\
 To be published in J. of Phys: A Math. and Theor.}

 \author{D.\ J.\ Rowe}
 \affiliation{
 Department of Physics, University of Toronto, Toronto, ON M5S 1A7, Canada}
  \date{December 2014}

 \begin{abstract}  

Capelli identities are shown to facilitate the construction of representations of  various Heisenberg algebras that arise in many-particle quantum mechanics and  the construction of holomorphic representations of many Lie algebras by Vector Coherent State methods.
We consider the original Capelli identity and its  generalizations by Turnbull and by Howe and Umeda.

 \end{abstract}
 \maketitle
 


 \section{Introduction}

The  Capelli identities  \cite{Capelli87,Weyl46,Turnbull48,HoweU91,KostantS91}
play central roles in invariant theory \cite{Weyl46,Howe89},
but have received little attention in physics.
However, as this paper shows, they are powerful tools for constructing asymptotic 
macroscopic limits of Lie algebra representations and boson approximations,  
of importance in physics, and  they provide the essential information 
needed to complete the VCS (vector-coherent-state) construction 
\cite{RoweRC84,Rowe84,RoweRG85,Hecht87,RoweR91,Rowe12} of induced holomorphic representations of Lie algebras.
Note that, by constructing a representation of a Lie algebra we mean  more than simply defining a module for the representation.  
A construction includes specifying a basis for the representation, which for a unitary representation should be orthonormal, and showing how it transforms under the action of Lie algebra elements.  
For the purposes of this paper,  basis states  are needed which separate into subsets that transform irreducibly under the actions of subalgebras and other Lie algebras.

This paper will show that a Lie algebra with holomorphic representations can be contracted to a  semi-direct sum of a compact subalgebra and one of the
Heisenberg algebras underlying the Capelli identities.
For each irrep (irreducible representation) of the compact  subalgebra, a corresponding unitary holomorphic irrep is  constructed for the contracted Lie algebra.
A motivation for this construction is that many of the spectrum generating algebras for nuclear models belong to the class of Lie algebras considered and their contractions are realised in macroscopic limits.
A more compelling motivation, is that the holomorphic representations of these semi-direct sum Lie algebras provide the necessary ingredients for the  VCS construction of the  holomorphic representations of the original uncontracted Lie algebra which are needed, for example, in applications to finite nuclei.
This will be shown in a following publication.

 The three standard Capelli identities are  expressed in terms of complex variables 
 $ \{ z_{ij}, i,j = 1,\dots, n\}$  and derivative operators 
$\{\partial_{ij} :=\partial/\partial z_{ij}\}$.
In the original Type I  identity \cite{Capelli87,Weyl46}, the $z_{ij}$ variables are independent and, together with the derivative operators,  satisfy the commutation relations
\beq [\partial_{ij} , z_{kl} ] = \delta_{i,k} \delta_{j,l} \eeq
of a Heisenberg algebra.
In the Type II generalization by Turnbull \cite{Turnbull48}, the variables are symmetric ($z_{ij} = z_{ji}$) and,  with the symmetrized derivative operators 
$\partial_{ij} := (1+ \delta_{i,j}) \partial/\partial z_{ij}$,  satisfy the commutation relations
\beq [\partial_{ij}, z_{kl} ] =\delta_{i,k}\delta_{j,l}  + \delta_{j,k}\delta_{i,l},\eeq
which are likewise those of a Heisenberg algebra.
In the Type III generalization by Howe and Umeda \cite{HoweU91} (see also Kostant and Sahi \cite{KostantS91}),
the variables are anti-symmetric ($z_{ij} = -z_{ji}$) and,  with the derivative operators $\partial_{ij} =\partial/\partial z_{ij}$,  satisfy the commutation relations
\beq [\partial_{ij}, z_{kl} ] =\delta_{i,k}\delta_{j,l} -\delta_{j,k}\delta_{i,l} ,\eeq
which are again those of a Heisenberg algebra.

Recall that the Heisenberg algebra of complex variables $\{z_i, i=1,\dots, n\}$ and  derivative operators $\{\partial/\partial z_{i}\}$, with commutation relations
\beq [\partial_{i} , z_j ] = \delta_{i,j}, \eeq
is realised in a  Bargmann representation \cite{Bargmann61} in which the raising and lowering operators 
$\{c^\dag_i, i = 1, \dots ,n\}$ and  $\{c_i, i = 1, \dots ,n\}$,  
 of an $n$-dimensional  oscillator are represented
 \beq c_i^\dag \to z_i , \quad c_i \to \partial_i =\partial / \partial z_i , 
 \quad i = 1, \dots n,\eeq
 on a Hilbert space $\Bb^{(n)}$ of entire analytic functions of the $\{ z_i\}$  variables.
Such a Bargmann Hilbert space  is  spanned by monomials  
$\psi_p = \prod_{i=1}^n z_i^{p_i}$ with inner products
\beq \langle \psi_q|\psi_p\rangle 
= \prod_{i=1}^n  \left.\partial_i^{q_i} z_i^{p_i} \right|_{z=0} 
= \delta_{q,p} \prod_{i=1}^n p_i!  .
\eeq
Thus, the Bargmann representation of a Heisenberg algebra with $n$ linearly-independent  $\{z_i\}$ elements is well understood in a monomial basis, and its Hilbert space naturally decomposes into subspaces that span irreps of 
$\mathfrak{gl}(n)$ and its $\mathfrak{u}(n)$ real form.
However, the Heisenberg algebras with $\{ z_{ij}\}$ variables having double indices have extra structure and their Bargmann spaces are Hilbert spaces for representations of different  Lie algebras.
Thus, the construction of the unitary irrep of a Heisenberg algebra in such variables,  in bases that belong to irreps of the different Lie algebras, is a non-trivial task which, as we show, is made possible by use of the Capelli identities.

The representation of a Heisenberg algebra with symmetric variables
$\{ z_{ij}= z_{ji}, i,j = 1,2,3\}$  relative to an orthonormal U$(3)$-coupled basis was  derived by Quesne (1981) \cite{Quesne81}, for use in interacting-boson models \cite{DzholosDJ75,ArimaI76,ArimaI78,IachelloA87}, and by Rosensteel and Rowe (1983) \cite{RosensteelR83} for use in a macroscopic approximation \cite{RosensteelR81,RoweR82} to the microscopic symplectic model of nuclear collective states \cite{RosensteelR77a,RosensteelR80}.
Following the introduction of VCS theory 
\cite{RoweRC84,Rowe84}, it became apparent that the unitary representations of the symmetric Heisenberg matrix algebras in  U$(3)$-coupled bases are needed for the VCS construction of  the holomorphic  representations of the non-compact Lie algebra 
$\mathfrak{sp}(3,\Rb)$ in related U$(n)$-coupled bases; these expressions of the holomorphic representations of $\mathfrak{sp}(3,\Rb)$ are central to the many-nucleon theory of nuclear collective dynamics \cite{Rowe85}.
In fact,  the representations of the Heisenberg matrix algebras associated with all the Capelli identities  enable the construction of orthonormal basis states for the holomorphic  representations of many Lie groups and 
their Lie algebras 
by algebraic  methods which avoid the difficult inner products of Harish-Chandra theory \cite{HarishChandra55IV,HarishChandra56V}  and the restriction of holomorphic representations to those of discrete series.
This subject is reviewed and developed in a following paper \cite{Rowe15} (see also sect.\ \ref{sect:conc}).
In consideration of these possibilities, it was recognised  \cite{LeBlancR87} that the needed  representations of the  Heisenberg algebras  follow from the Capelli identities.

This paper extends and generalises the results of ref.\ \cite{LeBlancR87}, which did not consider the Heisenberg algebra associated with the  Type I Capelli identity and was written before the Type III identity was discovered.
It shows that the Capelli identities lead to algorithms for the construction of irreducible unitary  representations of the corresponding Heisenberg algebras, albeit each in different but equally useful bases.

\section{Contractions of a Lie algebra}

A procedure for deriving  contractions of a Lie algebra has been  given by 
\.In\"on\"u and Wigner \cite{InonuW53} in which some elements of the Lie algebra contract to  an Abelian algebra. 
Other contractions have been defined, for example, by Saletan \cite{Saletan61},
 Weimer-Woods \cite{WeimerWoods91}, and others.
The following  definition is appropriate for the construction of holomorphic representations, in which subsets of elements contract  to Heisenberg algebras.  
Such contractions,  lead to so-called boson approximations and 
underlie  models in physics that become accurate in limiting 
(e.g., large particle number) situations  \cite{KleinM91,RosensteelR81}.  
They also provide  leading-order terms in the precise boson expansions of VCS theory and essential steps in the construction of  holomorphic representations of the Lie algebras to which they apply.

Let $\mathfrak{g}_0$ be  a real  semi-simple or reductive Lie algebra which,
as a vector space,  has a direct sum decomposition
\beq  \mathfrak{g}_0 = \mathfrak{h}_0 +  \mathfrak{p}_0, \eeq 
where $ \mathfrak{h}_0$ is a subalgebra and $\mathfrak{p}_0$ is invariant under the adjoint action of $\mathfrak{h}_0$ 
i.e., $[ \mathfrak{h}_0, \mathfrak{p}_0] \subset  \mathfrak{p}_0$. 
The complex extension $\mathfrak{g}_0$ of $\mathfrak{g}$ has the parallel  decomposition
\beq  \mathfrak{g} = \mathfrak{h} +  \mathfrak{p}. \eeq
A map  $\mathfrak{g}\to\bar{\mathfrak{g}} : X \mapsto \bar X$, from $\mathfrak{g}$  to a semi-direct sum Lie algebra $\bar{\mathfrak{g}}$, for which 
$\bar{\mathfrak{p}}$ is an ideal, is  said to be a contraction of $\mathfrak{g}$ if  it
maps $\mathfrak{h} \to \bar{\mathfrak{h}}$ and $\mathfrak{p} \to\bar{\mathfrak{p}}$ such that
\beqa
&{\rm (i)} \qquad& \text{$\mathfrak{h} \to \bar{\mathfrak{h}}$ is an isomorphism,} \label{eq:condo} \\
&{\rm (ii)}\qquad & [\bar X, \bar Y]= \bar Z 
                   \text{\;\;if\;\;} [X,Y]=Z , \quad \forall \; X\in\mathfrak{h}
                                                   \text{\;\;and\;\;} \forall \;  Y, Z \in\mathfrak{p}, \label{eq:cond2} \\
&{\rm (iii)} \qquad & [\bar Y, \bar Z]
                                                       \in \mathfrak{C},     
                            \quad \forall \;  Y, Z \in\mathfrak{p},   \label{eq:cond3}                          
\eeqa
where $\mathfrak{C}$  denotes the complex numbers regarded as a one-dimensional subalgebra of $\bar{\mathfrak{p}}$.
Condition (ii) ensures that the elements of $\bar{\mathfrak{p}}$ transform under the adjoint action of 
$\bar{\mathfrak{h}}$ in the same way as the elements of $\mathfrak{p}$ transform under the adjoint action of $\mathfrak{h}$.  
Condition (iii) is defined such that the elements of $\mathfrak{p}$ contract to elements of either an Abelian or (with the inclusion of an identity element in $\bar{\mathfrak{p}}$) a Heisenberg algebra.
Such contractions of $\mathfrak{g}$  define  corresponding contractions
 $\bar{\mathfrak{g}}_0\to\bar{\mathfrak{g}}_0$ of its real forms.
They are particularly useful  when $\mathfrak{h}_0$ is compact and
 $[\mathfrak{p},\mathfrak{p}] \subset \mathfrak{h}$.
If $G_0$ and $H_0$ are the maximal connected subgroups with Lie algebras given, respectively, by  $\mathfrak{g}_0$ and $\mathfrak{h}_0$, 
then the factor space $G_0/H_0$ is  a symmetric space
and VCS theory gives a prescription for 
inducing holomorphic irreps of $\mathfrak{g}_0$ and $G_0$ from irreps of 
$\mathfrak{h}_0$ and $H_0$ \cite{RoweR91, Rowe12}.

For example, $\mathfrak{su}$(3) has a direct sum vector-space decomposition
\beq \mathfrak{su}(3) =\mathfrak{so}(3) + \mathfrak{p}_0 , \eeq
in which $\mathfrak{p}_0$ is a 5-dimensional subset of elements of 
 $\mathfrak{su}(3)$ that transform as the components of an angular-momentum $L=2$ irrep under the adjoint action of  $\mathfrak{so}(3)$, i.e., 
 $[\mathfrak{so}(3), \mathfrak{p}_0] \subset \mathfrak{p}_0$,
  and 
 $[\mathfrak{p}_0, \mathfrak{p}_0] \subset \mathfrak{so}(3)$.
The $\mathfrak{su}(3)$ elements of this decomposition satisfy commutation relations
 \beqa   &&{[}L_i, L_j] \in \mathfrak{so}(3), \;\; 
 \forall\, L_i, L_j \in \mathfrak{so}(3) ,\\
 &&  {[}L_i, Q_{\nu}] \in \mathfrak{p}_0, \;\; \forall \, L_i \in \mathfrak{so}(3)
 \;\text{and}\; \forall \, Q_\nu\in \mathfrak{p}_0 ,  \\
&& {[}Q_\mu,Q_\nu{]} \in \mathfrak{so}(3), \;\; \forall\, Q_\mu, Q_\nu \in \mathfrak{p}_0 .
 \eeqa
There is then  a contraction of $\mathfrak{su}(3)$ to a semi-direct sum Lie algebra 
$\Rb^5 \sdsum \mathfrak{so}(3)$  in which the  $\mathfrak{so}(3)$ subalgebra is unchanged and the contracted elements 
of $\mathfrak{p}_0$ continue to transform as components of an $L=2$ tensor under the adjoint action of $\mathfrak{so}(3)$ but span a real five-dimensional Abelian Lie algebra isomorphic to $\Rb^5$.
This contraction is of  interest because it exposes the macroscopic limit of 
the  angular-momentum states of a large-dimensional SU(3) irrep to be those of a rigid-rotor  \cite{Ui70}
in which the intrinsic states of the rotor are described by a representation of the
$\Rb^5$ subalgebra and its rotational states carry representations of 
$\mathfrak{so}(3)$.
The following examples,  in which $\bar{\mathfrak{p}}$ includes an identity and is a Heisenberg algebra,  are of relevance to this paper

\subsection{Contraction of  $\mathfrak{u}(p+q)$ and $\mathfrak{u}(p,q)$ Lie algebras}
\label{sect:u(p,q)}

As a first example, consider the Lie algebras 
$\mathfrak{u}(p+q)$ and $\mathfrak{u}(p,q)$ with a common compact
subalgebra $\mathfrak{h}_0 = \mathfrak{u}(p)\oplus\mathfrak{u}(q)$
and common complex extension $\mathfrak{g} = \mathfrak{gl}(p+q)$.
This  extension is a matrix group with elements
\beq
\left(\begin{matrix} X&A\\ B&Y\end{matrix}\right) \quad \text{with} \quad
\begin{array} {ccc}
X\in \mathfrak{gl}(p), & \quad & A \in \mathfrak{p}_+, \\  
B\in \mathfrak{p}_- , & \quad & Y \in \mathfrak{gl}(q),
\end{array} 
\eeq
where $\mathfrak{p}_\pm$ are, respectively, Abelian subalgebras of raising and lowering operators and  $\mathfrak{p}$ is their vector-space direct sum
$\mathfrak{p}= \mathfrak{p}_+ + \mathfrak{p}_-$
with $[\mathfrak{p}_-, \mathfrak{p}_+]\in \mathfrak{h}$.
An equivalent realisation is given in differential form by
\beqa
X = \sum_{i=1}^q\sum_{j=1}^q X_{ij} x_i \frac{\partial}{\partial x_j}, \; 
&\qquad & 
A = \sum_{i=1}^q \sum_{\alpha=1}^p
A_{i\alpha} x_i \frac{\partial}{\partial y_\alpha}, \\
 B = \sum_{\alpha=1}^p \sum_{i=1}^q
 B_{\alpha i}  y_\alpha \frac{\partial}{\partial x_i},   
&\qquad & 
Y = \sum_{\alpha=1}^p \sum_{\beta=1}^p 
Y_{\alpha\beta} y_\alpha \frac{\partial}{\partial y_\beta} .
\eeqa
It is now seen that there are contractions in which
\beqa 
X \to \mathbb{X} +
\sum_{ij\alpha} X_{ij}  z_{i\alpha}\partial_{j\alpha} ,
&\qquad & 
\quad A \to k\sum_{i\alpha} A_{i\alpha} z_{i\alpha} ,\\
 B \to k^*\sum_{\alpha i} B_{\alpha i} \partial_{i\alpha} ,  \quad
&\qquad & 
Y \to \mathbb{Y} - \sum_{\alpha\beta i}
Y_{\alpha\beta} z_{i\beta} \partial_{i\alpha} ,
\eeqa
where $\partial_{i \alpha} = \partial /\partial z_{i\alpha}$, and
$\mathbb{X}$ and $\mathbb{Y}$ are representations of the respective elements  
$X\in\mathfrak{u}(p)$ and $Y\in\mathfrak{u}(q)$  that commute with the 
$\{ z_{i\alpha}\}$ variables and their derivatives.
In accordance with the above definition, this contraction conserves all the commutation relations of the $\mathfrak{gl}(p+q)$ Lie algebra except for those of  $A\in \mathfrak{p}_+$ and $B\in \mathfrak{p}_-$ for which 
\beq [B,A] \to |k|^2 \sum_{i\alpha} B_{\alpha i} A_{i\alpha} \in \mathfrak{C} .\eeq
The constant $k$ in this contraction is, in principle, arbitrary.
But, specific and possibly $i\alpha$-dependent values
will be appropriate for different irreps as they approach asymptotic limits.
These contractions  define a semi-direct sum of 
$\mathfrak{u}(p)\oplus\mathfrak{u}(q)$ and a Heisenberg Lie algebra.

 It will be shown in the following that, with a knowledge of elementary
 $\mathfrak{u}(p)$ and $\mathfrak{u}(q)$  vector-coupling coefficients (also called Clebsch-Gordan and Wigner coefficients),
the unitary irrep of the  Heisenberg algebra that emerges on the space of polynomials in the above $\{ z_{i\alpha}\}$ variables,  
and corresponding irreps of the $\mathfrak{u}(p+q)$ and $\mathfrak{u}(p,q)$ contractions, can be constructed in a 
$\mathfrak{u}(p)\+\mathfrak{u}(q)$ basis by use of the first Capelli identity.

This $\mathfrak{g} \to \bar{\mathfrak{g}}$ contraction can be used to derive an important approximation in physics,  known as the RPA (random phase approximation)
\cite{BohmP53}, which has many applications in the quantum theory of many-particle systems; see ref.\ \cite{Rowebook70} for a review and an algebraic expression of the RPA 
in a nuclear physics context.  
The RPA is  a quantal version of the classical theory of small-amplitude normal-mode vibrations.
Let $\mathfrak{g}_0 = \mathfrak{u}(p+q)$ be the unitary transformations of a $(p+q)$-dimensional Hilbert space  $\scrH$ of single-fermion states.
A Hilbert space $\Hb$ for $q$ fermions is then defined as the space spanned by all exterior products of $q$  states in $\scrH$.
A Hamiltonian $  H$ for this space, with interactions between pairs of fermions, is a Hermitian combination of linear and quadratic elements of 
$\mathfrak{g}= \mathfrak{gl}(p+q)$.
A  step  towards determining   low-energy states of  $  H$  is to separate the Hilbert space $\scrH$ into  subspaces
\beq \scrH = \scrH^{(q)} \+ \scrH^{(p)} ,\eeq
such that the energy $\langle\phi |  H \phi\rangle$ of the $q$-fermion state $|\phi\rangle$, given by the exterior product of an orthonormal basis for $\scrH^{(q)}$, minimises the energy among all such states for different choices of $\scrH^{(q)}$.
Such a minimum-energy state $|\phi\rangle$ defines a zero-order Hartree-Fock approximation for the ground state of the Hamiltonian $  H$.   It also defines a decomposition of the Lie algebra $\mathfrak{g}= \mathfrak{gl}(p+q)$ into two parts, 
$\mathfrak{h} + \mathfrak{p}$ as defined above, 
with $\mathfrak{h} = \mathfrak{gl}(q) \+ \mathfrak{gl}(p)$.
It is then shown by Hartree-Fock variational methods 
(cf.\ \cite{RoweWood10}, for example)
that, because of the separation of  $\scrH$ and $\mathfrak{h}$ into complementary subspaces  by minimisation of 
$\langle\phi |  H \phi\rangle$, the Hamiltonian $  H$ is expressible as a sum of linear terms in $\mathfrak{h}$ and interaction terms that are bilinear in the elements of $\mathfrak{p}$.  
It is also determined that, for many-particles (i.e., large values of $q$ and $p$), there is a neighbourhood of the Hilbert space 
$\Hb$ about the state $|\phi\rangle$ on which the action of the Lie algebra $\mathfrak{g}$ is accurately replaced by its contraction limit.  
Moreover, the Hamiltonian on this neighbourhood  becomes a quadratic function of the elements of the Heisenberg algebra of the coupled harmonic oscillator form
\beq   H \to   H_{\rm RPA}=
E_0 + \sum_{\mu\nu i j} V_{\mu i \nu j} z_{i\mu} \frac{\partial}{ \partial z_{j\nu}}
+\sum_{\mu\nu i j} W_{\mu i \nu j} z_{i\mu}  z_{j\nu}
+\sum_{\mu\nu i j} W^*_{\mu i \nu j} \frac{\partial^2}{\partial z_{i\mu}\partial z_{j\nu}},
\eeq
and easily diagonalised to produce an improved approximation for a so-called correlated ground state and its low-energy excitations.
The RPA proves to be a good approximation in many systems for which the ground state is already reasonably-well approximated by  $|\phi\rangle$.  
One can now anticipate that with the RPA set on a clear algebraic foundation, higher-order approximations to it will emerge naturally from the holomorphic representations  of $\mathfrak{u}(p+q)$ in a $\mathfrak{u}(q) \+ \mathfrak{gl}(p)$ basis.

\subsection{Contraction of $\mathfrak{sp}(N)$ and $\mathfrak{sp}(N,\Rb)$ Lie algebras}

The compact $\mathfrak{sp}(N)$ and non-compact $\mathfrak{sp}(N,\Rb)$ Lie algebras have a common complex extension $\mathfrak{sp}(N,\mathfrak{C})$ 
and a common compact subalgebra  $\mathfrak{u}(N)$ for which  
$\mathfrak{gl}(N)$ is the complex extension.
The algebra $\mathfrak{sp}(N,\mathfrak{C})$ has a simple realization with basis given in terms of variables $\{ x_i, i= 1, \dots, N\}$ and corresponding derivative operators by
\beq C_{ij} 
=\frac12 \left( x_i \frac{\partial}{\partial x_j }+  \frac{\partial}{\partial x_j}  x_i\right),
\quad A_{ij} = x_i x_j \quad B_{ij} = \frac{\partial^2}{\partial x_i \partial x_j} ,
\eeq
It also has a contraction 
in which
\beq C_{ij} \to \Cb_{ij} +  \sum_s   z_{is} \partial_{js} , \quad     
\quad A_{ij}  \to k z_{ij}, \quad B_{ij} \to k^* \partial_{ij} , \label{eq: contractedSp}
\eeq
for some $k\in \mathfrak{C}$, where the elements $\{ \Cb^{(p)}_{ij}\}$ are  representations of  elements of 
$\mathfrak{gl}(n)$ that commute with the $\{ z_{ij}\}$ variables and their derivatives, and
\beq z_{ij}= z_{ji}, \quad \partial_{ij} = (1+\delta_{i,j})\partial/\partial z_{ij}.
 \eeq
Thus, $\{ z_{ij}\}$ and $\{ \partial_{ij}\}$ are  raising and lowering operators for a Heisenberg algebra of the Type II Capelli identity
and the above contractions are  representations of a semi-direct sum of 
$\mathfrak{u}(N)$ and such a Heisenberg Lie algebra.

The non-compact $\mathfrak{sp}(N,\Rb)$ algebras and the Lie groups they generate have important applications in physics, e.q., in beam optics for particles and light \cite{GuilleminS77,Dragt87} and in the microscopic theory of nuclear collective dynamics \cite{RosensteelR77a,Rowe85,DytrychSBDV08a}. 
The compact $\mathfrak{sp}(N)$ symplectic algebras and associated Lie groups are also used extensively in the coupling schemes of  atomic \cite{Racah51} and nuclear \cite{Flowers52,French60} shell models.
The contraction limits of the symplectic algebras  characterize the  quantum dynamics of these systems as they approach their  macroscopic limits.
Particularly importantly, is the fact that uncontracted representations of both the compact and non-compact symplectic Lie algebras and groups can be induced, by VCS methods,  in a computationally tractable manner from the representations of their contractions.

\subsection{Contraction of $\mathfrak{so}(2N)$ and $\mathfrak{so}(2N){^*}$ Lie algebras}

The  $\mathfrak{so}(2N)$ and $\mathfrak{so}(2N){^*}$ Lie algebras 
have a common complex extension $\mathfrak{so}(2N,\mathfrak{C})$ 
and a common compact subalgebra given again by $\mathfrak{u}(n)$ 
with complex extension $\mathfrak{gl}(n)$.
The Lie algebra  $\mathfrak{so}(2n,\mathfrak{C})$ has a simple realisation with basis elements
\beq C_{ij} = \tfrac12 (a^\dag_i a_j-a_ja^\dag_i) , 
\quad A_{ij} =  a^\dag_i a^\dag_j, \quad B_{ij} =a_j a_i ,  \label{eq:Fpairalg}
\eeq
 expressed in terms of fermion  creation and annihilation operators
 $\{a^\dag_i , i=1, \dots, n\}$ and $\{ a_i, i= 1, \dots, n\}$  that obey anti-commutation relations
\beq \{ a_i, a^\dag_j\} = \delta_{i,j}, \quad \{ a_i, a_j\} = \{ a^\dag_i, a^\dag_j\} = 0 ,
\eeq
where the anti-commutator of two operators $X$ and $Y$ is defined by
$ \{ X,Y\} = XY + YX$.

For a complex  parameter $k$, the  
$\mathfrak{so}(2N,\mathfrak{C})$ algebra  has a contraction
\beq C_{ij} \to \Cb_{ij} + \sum_s z_{is}\partial_{js} , \quad
 A_{ij}  \to  k  z_{ij} \quad B_{ij} \to k^* \partial_{ij} ,
\eeq
where $ \Cb_{ij}$   represents an  element of $\mathfrak{gl}(n)$ that 
commutes with the $\{ z_{ij}\}$ variables and their derivatives, and
\beq z_{ij}= -z_{ji}, \quad \partial_{ij} = \partial/\partial z_{ij}.   
 \eeq
The $\{ z_{ij}\}$ variables and the derivative operators $\{ \partial_{ij}\}$ are now the raising and lowering operators of a Heisenberg algebra of the Type III Capelli identity.
These contractions are again seen to be representations of a semi-direct sum of 
$\mathfrak{u}(N)$ and a Heisenberg Lie algebra.

The compact SO$(2N)$ Lie group is of significance in the physics of fermion superconductivity in which it features as the group of Bogolyubov-Valatin transformations that leave the fermion anti-commutation relations invariant \cite{RoweWood10,RoweCR12}. 
Thus, this group SO$(2N)$ as well as it extension to 
SO$(2N+1)$ feature in theories of fermion pair coupling (cf., for example, ref.\
\cite{NishiyamaF92} and references therein). 
The contractions of the $\mathfrak{so}(2n)$ Lie algebra, known as the fermion pair algebra,  are  relevant to many-fermion models in which pairs of fermions are treated as bosons \cite{RoweC86,KleinM91}.


\section{Representations of the three classes of 
Heisenberg algebra}

The contraction of a Lie algebra $\mathfrak{g}= \mathfrak{h} +  \mathfrak{p}$ is defined above in terms of two algebras: $\mathfrak{h}$ and a Heisenberg algebra.  
A Heisenberg algebra has a single unitary irrep to within unitary equivalence \cite{Stone32,vonNeumann32}.
However, to construct a unitary irrep of the contracted Lie algebra, the irrep of the Heisenberg algebra is needed in a basis 
that separates its Hilbert space into irreducible subspaces of $\mathfrak{h}$.
Such a basis is said to reduce the representation of 
$\mathfrak{h}$ on the given Hilbert space.

\subsection{Heisenberg algebras related to type I Capelli algebras}
\label{sect:Capell1rep}

The Heisenberg algebra that  emerges in the contraction of the general linear algebra $\mathfrak{gl}(p+q)$  contains a  set of variables
$\{ z_{i\alpha}; i=1, \dots, p, \alpha = 1, \dots, q\}$ and their derivatives 
$\{ \partial_{i\alpha} :=\partial/\partial z_{i\alpha}\}$ that satisfy the commutation relations
\beq [\partial_{i\alpha} ,z_{j\beta}] = \delta_{i,j} \delta_{\alpha,\beta} . \eeq
General linear groups GL$(p)$ and GL$(q)$ have fundamental representations as left and right linear transformations, respectively, of the $\{ z_{i\alpha}\}$ variables.  
Elements of their $\mathfrak{gl}(p)$ and
$\mathfrak{gl}(q)$ Lie algebras then have corresponding realisations given by
\beq  L^{(p)}_{ij} := \sum_{\alpha=1}^q z_{i\alpha} \partial_{j\alpha} , 
 \quad  R^{(q)}_{\alpha\beta}  :=  - \sum_{i=1}^p z_{i\beta} \partial_{i\alpha} .
\eeq
Thus, in addition to carrying an irrep of a Heisenberg algebra, the Bargmann Hilbert space 
$\Bb^{(pq)}$ spanned by polynomials in the $\{ z_{i\alpha}\}$  variables  
is a module for a representation of the direct product  
${\rm GL}(p)\times{\rm GL}(q)$.
This representation is  a sum of outer tensor product irreps
$\bigoplus_\nu \{\nu\} \otimes \{\nu\}$, with highest weights 
\beq \blue \nu  = \black \{ \nu_1, \nu_2, \dots ,\nu_N\}, \quad
 \nu_1\geq \nu_2 \geq \dots \geq \nu_N \geq 0, \eeq
where $N= \min(p,q)$.
This decomposition of the representation of  ${\rm GL}(p)\times{\rm GL}(q)$ 
on the Bargmann space $\Bb^{(pq)}$ follows directly from the
Schur-Weyl duality theorem \cite{Weyl46}  and is a prototype of Howe duality
\cite{Howe85} (see also ref.\ \cite{RoweCR12} for a review of duality relationships in a physics context).
With an inner product for $\Bb^{(pq)}$ defined such that $\partial_{i\alpha}$ is the Hermitian adjoint of 
$z_{i\alpha}$, which we denote by $\partial_{i\alpha} = z_{i\alpha}^\dag$,  the representation of 
${\rm GL}(p)\times{\rm GL}(q)$ on $\Bb^{(pq)}$ restricts to a unitary representation of its
${\rm U}(p)\times{\rm U}(q)$ real form.

\subsubsection{Extremal states for  irreps of ${\rm U}(p)\times{\rm U}(q)$ on  subspaces of $\Bb^{(pq)}$}

Irreps of  ${\rm U}(p)\times{\rm U}(q)$  on subspaces of $\Bb^{(pq)}$ are conveniently characterised by extremal states, defined as states of highest weight relative to U$(p)$ and of lowest weight relative to U$(q)$. 
For example, the elementary polynomial $z_{11}$ is of highest U$(p)$ weight 
$\{ 1\} \equiv \{1,0, \dots \}$ and lowest 
U$(q)$ weight $\{ -1\} \equiv \{ -1,0, \dots \}$, and the polynomial 
$(z_{11} z_{22} - z_{12} z_{21})$  is of highest U$(p)$ weight $\{ 1^2\} \equiv \{1,1, 0, \dots )\}$ 
and lowest U$(q)$ weight $\{-1^2\} \equiv \{-1,-1, 0, \dots \} $.

Let $x_n$ and $\nabla_n$ denote the determinants
\beq 
x_n := \det \left( \begin{array}{cccc}
z_{11}& z_{12} & \dots & z_{1n} \\
z_{21}& z_{22} & \dots & z_{2n} \\
\vdots & \vdots & \ddots & \vdots \\
z_{n1}& z_{n2} & \dots & z_{nn} 
\end{array} \right) , \quad
\nabla_n := \det \left( \begin{array}{cccc}
\partial_{11}& \partial_{12} & \dots & \partial_{1n} \\
\partial_{21}& \partial_{22} & \dots & \partial_{2n} \\
\vdots & \vdots & \ddots & \vdots \\
\partial_{n1}& \partial_{n2} & \dots & \partial_{nn} 
\end{array} \right) ,  \label{eq:xn&nablan1}
\eeq
for 
$n=1, \dots ,  {\rm min}(p,q)$.   
Then, $x_n$ is a polynomial of highest U$(p)$ weight $\{1^n\}$
and lowest U$(q)$ weight $ \{ -1^n\}$. 
A general (non-normalised) extremal state for a  ${\rm U}(p)\times{\rm U}(q)$ 
irrep on a subspace of $\Bb^{(pq)}$
with highest U$(p)$ weight $\nu$ and lowest U$(q)$ weight $-\nu$
 is  given by 
\beq \psi_\nu^{(N)}(z) := x_1^{\pi_1} x_2^{\pi_2} \dots x_N^{\pi_N} ,
\text{ with\;\;} \nu_i = \sum_{j=i}^N \pi_j \; \text{ and\;\;} N\leq {\rm min}(p,q).
 \label{eq:psi_nu1}\eeq
Thus,  every such  ${\rm U}(p)\times{\rm U}(q)$ extremal weight is uniquely defined by its U$(p)$ highest weight component.

\subsubsection{Normalisation of ${\rm U}(p)\times{\rm U}(q)$ extremal states}

The normalisations of the above-defined  extremal states  are determined by use of the type I Capelli identity.
This  identity states that the product of the determinants $x_n$ and $ \nabla_n$
 satisfies the expression
\beq x_n \nabla_n =
\det [E^{(n)}_{ij} +(n-i)\delta_{i,j}] , \label{eq:Capelli1a}
\eeq
where
\beq E^{(n)}_{ij} = \sum_{s=1}^n z_{is} \partial_{js} , \quad i,j = 1, \dots , n\leq N,
\label{eq:Eij(n)}
\eeq
is an element of a $\mathfrak{u}(n)$ Lie algebra
and it is understood that the determinant on the right side of (\ref{eq:Capelli1a})  is defined with products of elements from different columns that preserve their  left to right order.
An equivalent expression is given by
\beq \nabla_n  x_n =
\det [E^{(n)}_{ij} +(n+1-i)\delta_{i,j}] . \label{eq:Capelli1b}
\eeq

The squared norm  of the wave function (\ref{eq:psi_nu1}) 
\beq  \langle  \psi^{(N)}_\nu | \psi^{(N)}_\nu\rangle =   
\langle \nabla_1^{\pi_1}\nabla_2^{\pi_2} \dots 
\nabla_{N-1}^{\pi_{N-1}}  \nabla_N^{\pi_N}
 x_N^{\pi_N}   x_{N-1}^{\pi_{N-1}}  \dots\, x_2^{\pi_2}  x_1^{\pi_1} \rangle 
\label{eq:normdefn} \eeq
is determined from the observation that, 
because an extremal  state $\psi$ satisfies the equation 
$E^{(N)}_{ij} \psi = 0$ for $i<j$, it is an eigenstate of the operator 
$\nabla_n x_n$ of eqn.\ (\ref{eq:Capelli1b}) with eigenvalue given by
\black
\beq \nabla_n x_n \psi = (E^{(n)}_{11}+n)(E^{(n)}_{22} + n-1) \dots (E^{(n)}_{nn}+1) \psi .
\label{eq:nablax_n1}
\eeq
Also, because $x_n$ is itself of extremal-weight  $\{ 1^n\}$, it follows that
\beqa \nabla^2_n x_n \psi &=&
\nabla_n (E^{(n)}_{11}+n)(E^{(n)}_{22} + n-1) \dots (E^{(n)}_{nn}+1) x_n \psi \\
&=&\nabla_n x_n  (E^{(n)}_{11}+n+1)(E^{(n)}_{22} + n) \dots (E^{(n)}_{nn}+2)  \psi \\
&=& \frac{(E^{(n)}_{11}+n+1)!\, (E^{(n)}_{22} + n)! \, \dots (E^{(n)}_{nn}+2)!}
{(E^{(n)}_{11}+n-1)! \, (E^{(n)}_{22} + n-2)! \, \dots E^{(n)}_{nn}!} \psi .
\eeqa
Proceeding in this way, it is determined that
\beq \nabla^\pi_n x^\pi_n \psi =
\frac{(E^{(n)}_{11}+n+\pi -1)!\,  (E^{(n)}_{22} + n+\pi -2)!\,  \dots (E^{(n)}_{nn}+\pi)!}
{(E^{(n)}_{11}+n-1)!\,  (E^{(n)}_{22} + n-2)! \, \dots E^{(n)}_{nn}!} \psi . \label{eq:nabla^px^p}
\eeq
The definition (\ref{eq:Eij(n)})  of $E^{(n)}_{ij}$, implies that
\beq E^{(n)}_{ii} \psi_\mu^{(m)} = E^{(m)}_{ii} \psi_\mu^{(m)}, \quad \text{for $n\geq m$}.
\eeq
It  follows that
\bal
& \langle \nabla_1^{\pi_1}  x_1^{\pi_1} \rangle =
\large\la\frac{(E^{(1)}_{11}+\pi_1)!}{E^{(1)}_{11} !}\large\ra = \pi_1! , \\
&\langle \nabla_1^{\pi_1} \nabla_2^{\pi_2} x_2^{\pi_2} x_1^{\pi_1} \rangle
= \large\la  \nabla_1^{p_1} 
\frac{(E^{(2)}_{11}+\pi_2+1)!\,(E^{(2)}_{22} + \pi_2)!}{(E^{(2)}_{11} +1)!\, E^{(2)}_{22}!}
  x_1^{\pi_1} \large\ra
= \frac{(\pi_1+\pi_2+1)! \,\pi_2!}{\pi_1+1} ,
\end{align}
and that
\bal
&\langle \nabla_1^{\pi_1} \nabla_2^{\pi_2} \nabla_3^{\pi_3} x_3^{\pi_3} x_2^{\pi_2} x_1^{\pi_1} \rangle
=\frac{(\pi_1+\pi_2+\pi_3+2)!\,(\pi_2+\pi_3+1)!\, \pi_3!}
{(\pi_1+1)(\pi_1+\pi_2+2) (\pi_2+1)}  .
\end{align}
Thus, in terms of weight components 
$\nu_i = \sum_{j=i}^n \pi_j$,
\bal
& \langle \psi^{(1)}_\nu |  \psi^{(1)}_\nu\rangle = \nu_1! \, , \\
& \langle \psi^{(2)}_\nu |  \psi^{(2)}_\nu\rangle =
\frac{(\nu_1+1)!\,\nu_2!} {\nu_1-\nu_2+1}\, ,\\
& \langle \psi^{(3)}_\nu |  \psi^{(3)}_\nu\rangle =
\frac{(\nu_1+2)!\,(\nu_2+1)!\, \nu_3!} 
{(\nu_1-\nu_2+1) (\nu_1-\nu_3+2)(\nu_2-\nu_3+1)}\, ,
\end{align}
and the pattern becomes recognizable.\\

\noindent {\bf Claim I:}
 The squared norm  of the wave function $\psi^{(N)}_\nu $ defined by eqn.\ (\ref{eq:psi_nu1}) is
 \beq {\cal N}^{(N)}_\nu = \langle  \psi^{(N)}_\nu | \psi^{(N)}_\nu\rangle =  
 \prod_{i=1}^{N} 
 \frac{(\nu_i+N-i)!}{\prod_{j=i+1}^N (\nu_i-\nu_j+j-i) }\, . \label{claim:I}
 \eeq
\\

\noindent {\bf Proof:}
The claim has been established  for $N\leq 3$.
For arbitrary $N$, the wave function  $\psi^{(N)}_\nu $ is a product
\beq  \psi^{(N)}_\nu = x_N^{\pi_N} \psi^{(N-1)}_\mu ,\eeq
where $\psi^{(N-1)}_\mu$ is a highest-weight polynomial of weight $\mu$ and
\beq \nu = (\mu_1+\pi_N, \mu_2+\pi_N, \dots ,\mu_{N-1}+\pi_N, \pi_N) .\eeq
Then
\beq {\cal N}^{(N)}_\nu = 
\langle  \psi^{(N-1)}_\mu | \nabla_N^{\pi_N} x_N^{\pi_N} \psi^{(N-1)}_\mu\rangle
\eeq
and,  by use of eqn.\ (\ref{eq:nabla^px^p}),
\beq {\cal N}^{(N)}_\nu = \frac{(\nu_1 + N-1)!\, (\nu_2 + N-2)! \dots \nu_N!}
{(\nu_1-\nu_N+N-1)! \,(\nu_2-\nu_N+ N-2)! \dots (\nu_{N-1}-\nu_N)!} 
\langle  \psi^{(N-1)}_\mu |\psi^{(N-1)}_\mu\rangle ,
\eeq
consistent with the claim.
Thus, because the claim is valid for $N\leq 3$, it is  valid for all $N$.
\hfill   $\square$

\subsubsection{Matrix elements of a type I Heisenberg algebra between
 ${\rm U}(p)\times{\rm U}(q)$ extremal states}

Given some elementary vector-coupling coefficients 
for the   $\mathfrak{u}(p)$ and $\mathfrak{u}(q)$ irreps, one can obtain all matrix elements of a type I Heisenberg algebra in a  ${\rm U}(p)\times{\rm U}(q)$-coupled basis
from the subset 
\beq \langle \nu + \Delta_k | z_{kk} |\nu\rangle 
= \langle \nu  | \partial_{kk} |\nu + \Delta_k\rangle^*,
\quad k = 1, \dots N = {\rm min}(p.q),  \label{eq:MEdeltakk}
\eeq
($*$ denotes complex conjugation) for normalised  extremal states
\beq
|\nu+\Delta_k\rangle := \frac{| \psi^{(N)}_{\nu + \Delta_k}\rangle}
{\left[ \langle \psi^{(N)}_{\nu + \Delta_k} | \psi^{(N)}_{\nu + \Delta_k}\rangle \right]^{\frac12}} , 
\quad 
 |\nu\rangle := \frac{| \psi^{(N)}_\nu\rangle}
{\left[ \langle \psi^{(N)}_{\nu } | \psi^{(N)}_{\nu}\rangle \right]^{\frac12}} , 
\eeq
where $\psi^{(N)}_\nu$ is defined by eqn.\ (\ref{eq:psi_nu1}) and
 $\Delta_k = \{0,\dots ,0, 1, 0, \dots,0\}$ is the $\mathfrak{u}(N)$ weight of 
$z_{kk}$ for which all but its $k$ component is equal to zero.
It follows from the expression (\ref{eq:psi_nu1}) of the wave function
$\psi^{(N)}_{\nu}$ that the corresponding
 wave function $\psi^{(N)}_{\nu + \Delta_k}$ with a shifted weight is  given by
\beq \psi^{(N)}_{\nu + \Delta_k} =  x_1^{\pi_1}\dots 
x_{k-2}^{\pi_{k-2}}  x_{k-1}^{\pi_{k-1}-1}x_{k}^{\pi_{k}+1} x_{k+1}^{\pi_{k+1}}
\dots x_N^{\pi_N} .
\eeq

Consider the ratio 
\bal R &:=  \frac{\langle \psi^{(N)}_{\nu + \Delta_k} | z_{kk} \psi^{(N)}_{\nu}\rangle}
{ \langle \psi^{(N)}_{\nu+ \Delta_k} |  \psi^{(N)}_{\nu+ \Delta_k} \rangle}
 = \frac{\langle \psi^{(N)}_{\nu } | \partial_{kk} \psi^{(N)}_{\nu + \Delta_k}\rangle}
{ \langle \psi^{(N)}_{\nu+ \Delta_k} |  \psi^{(N)}_{\nu+ \Delta_k} \rangle} \\
&=\frac{
\langle \nabla_1^{\pi_1} \dots \nabla_k^{\pi_k} \partial_{kk}   W 
x_k^{\pi_k+1} x_{k-1}^{\pi_{k-1}-1} x_{k-2}^{\pi_{k-2}} \dots x_1^{\pi_1} \rangle}
{\langle \nabla_1^{\pi_1} \dots \nabla_{k-2}^{\pi_{k-2}}  \nabla_{k-1}^{\pi_{k-1}-1}
\nabla_{k}^{\pi_{k}+1} W 
x_k^{\pi_k+1} x_{k-1}^{\pi_{k-1}-1} x_{k-2}^{\pi_{k-2}} \dots x_1^{\pi_1} \rangle},
\end{align}
with
\beq   W = \nabla_{k+1}^{\pi_{k+1}} \dots \nabla_N^{\pi_N} 
x_N^{\pi_N} \dots x_{k+1}^{\pi_{k+1}} .
\eeq
This ratio has the simplifying property that the function
$x_k^{\pi_k+1} x_{k-1}^{\pi_{k-1}-1} x_{k-2}^{\pi_{k-2}} \dots x_1^{\pi_1}$
 to the right of the operator $W$ is the same in both its numerator and denominator.
 Moreover, this function is an eigenfunction of $ W$.
Thus, its eigenvalue can be factored out  to give
\beq R= \frac{
\langle \nabla_1^{\pi_1} \dots \nabla_k^{\pi_k} [\partial_{kk} ,
x_k^{\pi_k+1}] x_{k-1}^{\pi_{k-1}-1} x_{k-2}^{\pi_{k-2}} \dots x_1^{\pi_1} \rangle}
{\langle \nabla_1^{\pi_1} \dots \nabla_{k-2}^{\pi_{k-2}}  \nabla_{k-1}^{\pi_{k-1}-1}
\nabla_{k}^{\pi_{k}+1}
x_k^{\pi_k+1} x_{k-1}^{\pi_{k-1}-1} x_{k-2}^{\pi_{k-2}} \dots x_1^{\pi_1} \rangle},
\eeq
where use is made of the observation that $[\partial_{kk} , x_{k'}]=0$ for all $k'<k$.
Now, with the  identity 
\beq [\partial_{kk}, x_k^{\pi_k+1}] = (\pi_k+1) x_{k-1} x_k^{\pi_k} , \eeq
it is determined that
\beqa R &=& (\pi_k+1) \frac{
\langle \nabla_1^{\pi_1} \dots \nabla_k^{\pi_k} 
x_k^{\pi_k} x_{k-1}^{\pi_{k-1}} x_{k-2}^{\pi_{k-2}} \dots x_1^{\pi_1} \rangle}
{\langle \nabla_1^{\pi_1} \dots \nabla_{k-2}^{\pi_{k-2}}  \nabla_{k-1}^{\pi_{k-1}-1}
\nabla_{k}^{\pi_{k}+1}
x_k^{\pi_k+1} x_{k-1}^{\pi_{k-1}-1} x_{k-2}^{\pi_{k-2}} \dots x_1^{\pi_1} \rangle} 
\nonumber\\
&=&(\pi_k+1) \frac{\langle \psi^{(k)}_{\mu} |\psi^{(k)}_{\mu}\rangle}
{\langle \psi^{(k)}_{\mu+ \Delta_k} |\psi^{(k)}_{\mu+ \Delta_k}\rangle} ,
\eeqa
where
\beq \psi^{(k)}_{\mu} := x_1^{\pi_1} x_2^{\pi_2} \dots x_k^{\pi_k}, \quad
\mu_i  = \sum_{j=i}^k \pi_j = \nu_i-\nu_{k+1}, \;\; i=1,\dots, k.\eeq
It follows that the desired matrix elements
\beq \langle \nu+\Delta_k| z_{kk} |\nu\rangle = \frac
{\langle \psi^{(N)}_{\nu + \Delta_k} | z_{kk} \psi^{(N)}_\nu\rangle}
{\left[ \langle \psi^{(N)}_{\nu + \Delta_k} | \psi^{(N)}_{\nu + \Delta_k}\rangle
\langle \psi^{(N)}_{\nu } | \psi^{(N)}_{\nu}\rangle \right]^{\frac12}} ,
\eeq
are given by
\beq \langle \nu+\Delta_k| z_{kk} |\nu\rangle = (\nu_k-\nu_{k+1}+1)
\frac{\langle \psi^{(k)}_\mu |\psi^{(k)}_\mu\rangle }
{\langle \psi^{(k)}_{\mu+\Delta_k} |\psi^{(k)}_{\mu+\Delta_k}\rangle}
\left[ \frac{\langle \psi^{(N)}_{\nu+ \Delta_k } | \psi^{(N)}_{\nu + \Delta_k}\rangle}
{\langle \psi^{(N)}_{\nu } | \psi^{(N)}_{\nu}\rangle}\right]^{\frac12} \label{eq:UpqMEs1}
\eeq
and are easily evaluated with the expression for the norms given by Claim I.

\subsubsection{Unitary representation of the type I Heisenberg algebra
in a ${\rm U}(p)\times{\rm U}(q)$-coupled  basis}

 The essential tool that makes it possible to determine
all matrix elements of the above-defined Heisenberg algebra 
in terms of elementary ${\rm U}(p)$ and ${\rm U}(q)$ 
vector-coupling coefficients is the Wigner-Eckart theorem 
 \footnote{In its original application to SU(2), the Wigner-Eckart theorem 
      \cite{Sakurai94} states that matrix elements of a spherical tensor 
     operator connecting the angular-momentum eigenstates of two SU(2) 
     irreps can be expressed as products of two factors, one of which is
     independent of angular momentum orientation, and the other is a 
     vector-coupling (Clebsch-Gordan) coefficient. 
     It is now known that, subject to a few conditions, the theorem also
     applies to the matrix elements of  other groups and Lie algebras
     (see for example \cite{Agrawala80}).
     In particular, it applies to the unitary groups.}.
     
The Wigner-Eckart theorem states, for example, that if $\lambda$ 
labels a U$(N)$ irrep and $\mu$ indexes an orthonormal basis for this irrep then, if $T_{\lambda_2}$ is a tensor operator whose components 
$\{T_{\lambda_2\mu_2}\}$ transform as basis states for a 
U$(N)$ irrep $\lambda_2$, the matrix elements of this tensor between  basis states of any two U$(N)$ irreps are  expressible as a sum of products
\beq \langle \lambda_3\mu_3 | T_{\lambda_2\mu_2} | \lambda_1\mu_1\rangle
=\sum_\rho (\lambda_1 \mu_1, \lambda_2\mu_2 |\rho \lambda_3\mu_3) \, 
\langle\lambda_3 \| T_{\lambda_2} \| \lambda_1\rangle_\rho , \label{eq:WEthm}
\eeq
where $\rho$  indexes the $n_{\lambda_3}$ occurrences 
of the U$(N)$ irrep $\lambda_3$ in the tensor product 
$\lambda_2\otimes \lambda_1= \bigoplus_\lambda n_\lambda \lambda$,
$(\lambda_1 \mu_1, \lambda_2\mu_2 | \rho\lambda_3\mu_3)$ is a vector-coupling coefficent, and the so-called reduced matrix element
$\langle\lambda_3 \| T_{\lambda_2} \| \lambda_2\rangle_\rho$ is independent of the choice of bases.
The theorem is valuable because the vector-coupling coefficients are determined by the properties of the U$(N)$ irreps, independently of  how they are realised in any particular situation.
Thus, if the vector-coupling coefficients are known, one has only to calculate $n_{\lambda_3}$ distinct  non-zero matrix element 
$\langle \lambda_3\mu_3 | T_{\lambda_2\mu_2} | \lambda_2\mu_2\rangle$ to
determine the reduced matrix elements 
$\langle\lambda_3 \| T_{\lambda_2} \| \lambda_2\rangle_\rho$.
From the unitarity of a U$(N)$ representation it also follows that the matrix elements of the Hermitian adjoints of a tensor operator are given by
\beq \langle \lambda_3\mu_3 | \big(T_{\lambda_2\mu_2}\big)^\dag
 | \lambda_1\mu_1\rangle
= \langle \lambda_1\mu_1 | T_{\lambda_2\mu_2} | \lambda_3\mu_3\rangle^*,
\label{eq:Hermadjs}
\eeq
where $*$ denotes complex conjugation.

Application of the Wigner-Eckart theorem to the calculation of matrix elements  of the Heisenberg algebras that we consider is simplified considerably by the fact that  the elements of these algebras are components of tensors $T_{\lambda_2}$ for which the tensor product $\lambda_2\otimes \lambda_1$ is always multiplicity free.  Thus, for present purposes, we have no need of the multiplicity index and all matrix elements of the Heisenberg algebra under consideration are defined by single matrix elements between extremal states.
A minor complication arises for the type I Heisenberg algebras because their elements  are components of a direct product ${\rm U}(p)\times {\rm U}(q)$ tensor.
Nevertheless, as we now show, the construction remains simple.

An element $z_{i\alpha}$ of the type I Heisenberg algebra is a component of  
weight $\Delta_i$ of a U$(p)$ tensor of  highest weight $\Delta_1$  
and a component of  weight $-\Delta_\alpha$ of a U$(q)$ tensor that is also of highest weight $\Delta_1$.
By definition  $|\nu\rangle$ is a state of highest  weight $\nu$ of a U$(p)$ irrep  and of lowest weight $-\nu$ of a U$(q)$ irrep of highest weight $\nu$.
Equation (\ref{eq:UpqMEs1}) gives a non-vanishing matrix element 
$\langle \nu' | z_{kk} |\nu\rangle$ with $k$ such that $\Delta_k = \nu' -\nu$.
Thus, according to the Wigner-Eckart theorem, the matrix element 
of eqn.\  (\ref{eq:UpqMEs1}) is  equal to the product
\beq \langle \nu' | z_{kk} |\nu\rangle = 
(\nu \nu , \Delta_1 \Delta_k | \nu'\nu') (\nu ,-\nu ,\, \Delta_1,- \Delta_k | \nu',-\nu')
\langle \nu'\| z \| \nu \rangle ,
\eeq
in which $(\nu \nu , \Delta_1 \Delta_k | \nu'\nu')$ 
and $(\nu ,-\nu ,\, \Delta_1,- \Delta_k | \nu',-\nu')$ are, respectively,
U$(p)$ and  U$(q)$ vector-coupling coefficients, and
$\langle \nu'\| z \| \nu \rangle$ is a ${\rm U}(p)\times{\rm U}(q)$-reduced matrix element for their combined irreps.
Thus, the matrix elements of any $z_{i\alpha}$ are obtained by the further use of eqn.\ (\ref{eq:WEthm})
and because $\partial_{i\alpha}$ is the Hermitian adjoint of $z_{i\alpha}$ in a unitary representation, its matrix elements  are also obtained from the Hermiticity relationship (\ref{eq:Hermadjs}).

The vector-coupling coefficients are well-known for U(2) and a computer code is freely available for those of U(3) \cite{AkiyamaD73}.
Restricted sets of coefficients are also available for other groups, such as U(4)
\cite{HechtPang69} and SO(5) \cite{CaprioRW09}
and algorithms have been developed
\cite{BiedenharnL68,FlathT93,RoweR95,RoweR96}
 for computing the coefficients of any unitary group.

\subsection{Heisenberg algebas related to type II Capelli algebras}
\label{sect:Capell2rep}

The Heisenberg algebra that  emerges in the contraction of the symplectic algebra 
$\mathfrak{sp}(N,\mathfrak{C})$ contains a  set of symmetric variables  and derivative operators
\beq  z_{ij}= z_{ji}, \quad \partial_{ij}= \partial_{ji} = (1+\delta_{i,j}) \partial/\partial z_{ij} ,
\quad i,j = 1, \dots, N ,\eeq
that satisfy the commutation relations
\beq [\partial_{ij}, z_{kl}] = \delta_{i,k}\delta_{j,l}+ \delta_{i,l}\delta_{j,k}.
\eeq
The Bargmann space $\Bb^{(\frac12 N(N+1)}$ spanned by polynomials in the
$\frac12 N(N+1)$ independent variables $\{ z_{ij}, i\leq j\}$ carries the 
unitary irrep of this Heisenberg algebra.
It is also a module for a reducible representation of the $\mathfrak{gl}(N)$ Lie algebra with elements
\beq E^{(N)}_{ij} = \sum_{s=1}^N z_{is} \partial_{js} .
\eeq
This representation is  a sum of GL$(N)$ irreps $\bigoplus_\nu  \{\nu\} $, with highest weights 
\beq \nu = \{ \nu_1, \nu_2, \dots ,\nu_N\}, \quad
 \nu_1\geq \nu_2 \geq \dots \geq \nu_N \geq 0, \eeq
in which, because $[E^{(N)}_{ii}, z_{kl}]= (\delta_{i,k} +\delta_{i,l})z_{kl}$
 implies that $z_{kl}$ is of weight $\Delta_k+\Delta_k$, 
each component $\nu_i$ is an even integer.
With an  inner product for $\Bb^{(\frac12 N(N+1))}$ defined such that 
$\partial_{ij}$ is the Hermitian adjoint of $z_{ij}$,   
the representation of ${\rm GL}(N)$ on $\Bb^{(\frac12 N(N+1))}$ 
restricts to a unitary representation of its
${\rm U}(N)$ real form.

\subsubsection{Highest-weight states for  irreps of ${\rm U}(N)$ on  subspaces of 
$\Bb^{(\frac12 N(N+1))}$}

The irreps of  ${\rm U}(N)$  on subspaces of $\Bb^{(\frac12 N(N+1))}$ are  characterised by highest-weight states, which are polynomials in the $\{ z_{ij}\}$  variables that are annihilated by the 
$\{ E^{(N)}_{ij}, i<j\}$ raising operators.
For example, the elementary polynomial $z_{11}$ is of highest U$(N)$ weight 
$\{2\} \equiv \{2,0, \dots \}$ and   the polynomial $(z_{11} z_{22} - z_{12} z_{21})$  is of highest U$(N)$ weight $\{2^2\} \equiv \{2,2, 0, \dots \}$.

In parallel with eqn.\ (\ref{eq:xn&nablan1}), let $x_n$ and $\nabla_n$ 
for $n = 1, \dots, N$ denote the determinants
\beq x_n := \det(z_{ij}) , \quad \nabla_n := \det(\partial_{ij}),  \quad
\text{with} \;\;  i,j = 1, \dots,n.
\label{eq:xn&nablan2} \eeq
Then, $x_n$ is a polynomial of highest U$(N)$ weight $\{ 2^n\}$.
A general  highest-weight state of weight $\nu$ for a  ${\rm U}(N)$ irrep on a subspace of  $\Bb^{(\frac12 N(N+1))}$ is  given by
\beq \psi_\nu^{(N)}(z) := x_1^{p_1} x_2^{p_2} \dots x_N^{p_N} 
\text{ with\;\;} \nu_i = \sum_{j=i}^N 2p_j .
 \label{eq:psi_nu2}\eeq

\subsubsection{Normalization factors for the  ${\rm U}(N)$ highest-weight states}

The normalizations of the above-defined  highest-weight states  are determined by use of the type II Capelli identity
\beq x_n \nabla_n = \det [E^{(n)}_{ij} +(n-i)\delta_{i,j}] , \label{eq:Capelli2a}
\eeq
in which
\beq E^{(n)}_{ij} = \sum_{s=1}^n z_{is} \partial_{js} , \quad i,j = 1, \dots , n\leq N,
\label{eq:Eij(n)2}
\eeq
is an element of a $\mathfrak{u}(n)$ Lie algebra
and it is understood that the determinant on the right side of (\ref{eq:Capelli2a})  is defined with products of elements from different columns that preserve their  left to right order.
An  equivalent expression is given by
\beq \nabla_n  x_n =
\det [E^{(n)}_{ij} +(n+2-i)\delta_{i,j}] . \label{eq:Capelli2b}
\eeq

The squared norm   
\beq  \langle  \psi^{(N)}_\nu | \psi^{(N)}_\nu\rangle =   
\langle \nabla_1^{p_1}\nabla_2^{p_2} \dots  \nabla_n^{p_N}
 x_n^{p_N}   \dots x_2^{p_2}  x_1^{p_1} \rangle ,
\label{eq:normdefn2} \eeq
of the wave function $\psi^{(N)}_\nu$ of eqn.\ (\ref{eq:psi_nu2})
is determined starting from the observation that, when acting on a highest-weight wave function $\psi$, the operator $\nabla_n x_n$ of eqn.\ (\ref{eq:Capelli2b}) simplifies to
\beq \nabla_n x_n \psi = (E^{(n)}_{11}+n+1)(E^{(n)}_{22} + n) \dots (E^{(n)}_{nn}+2) \psi .
\label{eq:nablax_n2}  \eeq
Proceeding as in sect.\ \ref{sect:Capell1rep}, it is  determined that
\beq \nabla^p_n x^p_n \psi =
\frac{(E^{(n)}_{11}+n+2p-1)!!\,  (E^{(n)}_{22} + n+2p-2)!!\,  \dots (E^{(n)}_{nn}+2p)!!}
{(E^{(n)}_{11}+n-1)!!\,  (E^{(n)}_{22} + n-2)!! \, \dots E^{(n)}_{nn}!!} \psi . \label{eq:nablax^p2}
\eeq
Similarly, with the recognition that
\beq E^{(n)}_{ii} \psi_\mu^{(m)} = E^{(m)}_{ii} \psi_\mu^{(m)}, \quad \text{if $n\geq m$},
\eeq
and evaluation of
$\langle \nabla_1^{p_1} \nabla_2^{p_2} x_2^{p_2} x_1^{p_1} \rangle$ and
$\langle  \nabla_1^{p_1} \nabla_2^{p_2} \nabla_3^{p_3} x_3^{p_3} x_2^{p_2} x_1^{p_1} \rangle $,
the pattern becomes recognizable.\\

\noindent {\bf Claim II:}
 The squared norm of the wave function $\psi^{(N)}_\nu $ of eqn.\ (\ref{eq:psi_nu2}) is 
 \beq {\cal N}^{(N)}_\nu = \langle  \psi^{(N)}_\nu | \psi^{(N)}_\nu\rangle =  
 \prod_{i=1}^{N} \left[ (\nu_i+N-i)!!
\prod_{j=i+1}^N
\frac{(\nu_i-\nu_j+j-i-1)!!}{(\nu_i-\nu_j+j-i)!!} \right]. \label{claim:II}
 \eeq

\noindent {\bf Proof:}
The wave function  $\psi^{(N)}_\nu $ for arbitrary $N$ is a product
\beq  \psi^{(N)}_\nu = x_N^{p_N} \psi^{(N-1)}_\mu ,\eeq
where $\psi^{(N-1)}_\mu$ is a highest-weight polynomial of weight $\mu$ and
\beq \nu = (\mu_1+2p_N, \dots ,\mu_{N-1}+2p_N, 2p_N) .\eeq
By use of eqn.\ (\ref{eq:nablax^p2}) and assuming the claim is correct for
${\cal N}^{(N-1)}_\mu$, it follows that
\beqa 
\langle  \psi^{(N-1)}_\mu | \nabla_N^{p_N} x_N^{p_N} \psi^{(N-1)}_\mu\rangle 
&=& \frac{(\nu_1 + N-1)! (\nu_2 + N-2)! \dots \nu_N!}
{(\nu_1-\nu_N+N-1)! (\nu_2-\nu_N+ N-2)! \dots (\nu_{N-1}-\nu_N)!} 
\langle  \psi^{(N-1)}_\mu |\psi^{(N-1)}_\mu\rangle \nonumber\\
&=& {\cal N}^{(N)}_\nu .
\eeqa
Thus, given that the claim is valid for $N\leq 3$, it is  valid for all $N$.
\hfill   $\square$

\subsubsection{Matrix elements of a type II Heisenberg algebra between 
${\rm U}(N)$ highest-weight states}

Given a knowledge of the irreps of  $\mathfrak{u}(N)$  and its coupling coefficients, one can obtain all matrix elements of the contracted
$\mathfrak{sp}(N,\mathfrak{C})$ Lie algebra from the matrix elements 
\beq \langle \nu + 2\Delta_k | z_{kk} |\nu\rangle 
= \langle \nu  | \partial_{kk} |\nu + 2\Delta_k\rangle^*, \quad k = 1, \dots N. \label{eq:MEdeltakk2}
\eeq
between normalized highest-weight states
\beq
|\nu+2\Delta_k\rangle := \frac{| \psi^{(N)}_{\nu + 2\Delta_k}\rangle}
{\left[ \langle \psi^{(N)}_{\nu + 2\Delta_k}|\psi^{(N)}_{\nu + 2\Delta_k}\rangle \right]^{\frac12}} , 
\quad 
 |\nu\rangle := \frac{| \psi^{(N)}_\nu\rangle}
{\left[ \langle \psi^{(N)}_{\nu } | \psi^{(N)}_{\nu}\rangle \right]^{\frac12}} , 
\eeq
where  $2\Delta_{k} = (0,\dots ,0, 2, 0, \dots,0)$ is the weight of $z_{kk}$  as a $\mathfrak{u}(N)$ raising operator with all but its $k$ component equal to zero.
Thus, with $\psi^{(N)}_\nu$ expressed 
as in eqn.\ (\ref{eq:psi_nu2}),
the shifted wave functions  take the form
\beq \psi^{(N)}_{\nu + 2\Delta_k} =  x_1^{p_1}\dots 
x_{k-2}^{p_{k-2}}  x_{k-1}^{p_{k-1}-1}x_{k}^{p_{k}+1} x_{k+1}^{p_{k+1}}
\dots x_N^{p_N} .
\eeq

Consider first the ratio 
\bal R &:= \frac{ \langle \psi^{(N)}_{\nu + 2\Delta_k} 
| z_{kk} \psi^{(N)}_{\nu}\rangle}
{ \langle \psi^{(N)}_{\nu+ 2\Delta_k} |  \psi^{(N)}_{\nu+ 2\Delta_k} \rangle} 
= \frac{ \langle \psi^{(N)}_{\nu} | \partial_{kk} \psi^{(N)}_{\nu+2\Delta_k}\rangle}
{ \langle \psi^{(N)}_{\nu+2\Delta_k} |  \psi^{(N)}_{\nu+2\Delta_k} \rangle} \\
&=\frac{
\langle \nabla_1^{p_1} \dots \nabla_k^{p_k} \partial_{kk}   W 
x_k^{p_k+1} x_{k-1}^{p_{k-1}-1} x_{k-2}^{p_{k-2}} \dots x_1^{p_1} \rangle}
{\langle \nabla_1^{p_1} \dots \nabla_{k-2}^{p_{k-2}}  \nabla_{k-1}^{p_{k-1}-1}
\nabla_{k}^{p_{k}+1}
   W 
x_k^{p_k+1} x_{k-1}^{p_{k-1}-1} x_{k-2}^{p_{k-2}} \dots x_1^{p_1} \rangle},
\end{align}
with
\beq   W = \nabla_{k+1}^{p_{k+1}} \dots \nabla_N^{p_N} 
x_N^{p_N} \dots x_{k+1}^{p_{k+1}} .
\eeq
The monomial to the right of the operator $  W$ is the same in both the numerator and denominator of $R$.  It is an eigenfunction of $  W$ and its eigenvalue can be factored out  to give
\beq R= \frac{\langle \nabla_1^{p_1} \dots \nabla_k^{p_k} [\partial_{kk} ,
x_k^{p_k+1}] x_{k-1}^{p_{k-1}-1} x_{k-2}^{p_{k-2}} \dots x_1^{p_1} \rangle}
{\langle \nabla_1^{p_1} \dots \nabla_{k-2}^{p_{k-2}}  \nabla_{k-1}^{p_{k-1}-1}
\nabla_{k}^{p_{k}+1}
x_k^{p_k+1} x_{k-1}^{p_{k-1}-1} x_{k-2}^{p_{k-2}} \dots x_1^{p_1} \rangle},
\eeq
where use is made of the observation that $[\partial_{kk} , x_{k'}]=0$ for all $k'<k$.
The identity 
\beq [\partial_{kk}, x_k^{p_k+1}] = 2(p_k+1) x_{k-1} x_k^{p_k} , \eeq
then leads to the expression
\beq R = 2 (p_k+1) 
\frac{\langle \psi^{(k)}_\mu |\psi^{(k)}_\mu\rangle }
{\langle \psi^{(k)}_{\mu+2\Delta_k} |\psi^{(k)}_{\mu+2\Delta_k}\rangle} ,
\eeq
where 
\beq \psi_\mu^{(k)} = x_i^{p_1} \dots x_k^{p_k}, \quad
\psi_{\mu+2\Delta_k}^{(k)} = 
x_i^{p_1} \dots x_{k-1}^{p_{k-1}-1} x_{k}^{p_{k}+1} ;
\eeq
which corresponds to setting
$\mu_i = \nu_i - \nu_{k+1}$, for $i=1,\dots, k$, and $2p_k = \nu_k - \nu_{k+1}$.

Finally, the desired matrix elements between highest-weight states are given by
\beq \langle \nu+2\Delta_k | z_{kk} |\nu\rangle = (\nu_k-\nu_{k+1}+2)
\frac{\langle \psi^{(k)}_\mu |\psi^{(k)}_\mu\rangle }
{\langle \psi^{(k)}_{\mu+2\Delta_k} |\psi^{(k)}_{\mu+2\Delta_k}\rangle}
\left[ \frac{\langle \psi^{(N)}_{\nu+ 2\Delta_k } | \psi^{(N)}_{\nu + 2\Delta_k}\rangle}
{\langle \psi^{(N)}_{\nu } | \psi^{(N)}_{\nu}\rangle}\right]^{\frac12} \label{eq:UNMEs}
\eeq
and are easily evaluated with the expression for the norms given by Claim II.

\subsubsection{Unitary representation of the type II Heisenberg algebra
in a ${\rm U}(N)$-coupled  basis}

An element $z_{kl}$ of the type II Heisenberg algebra  
is a component of  weight $\Delta_k +\Delta_l$ of a U$(N)$ tensor of  highest weight $2\Delta_1=(2, 0, \dots )$.
Equation (\ref{eq:UNMEs}) gives a non-vanishing matrix element 
$\langle \nu' | z_{kk} |\nu\rangle$ with $k$ such that $2\Delta_k = \nu' -\nu$.
Thus, according to the Wigner-Eckart theorem, the matrix element 
of eqn.\  (\ref{eq:UpqMEs1}) is  equal to the product
\beq \langle \nu' | z_{kk} |\nu\rangle = 
(\nu \nu ,\, 2\Delta_1, 2\Delta_k | \nu'\nu') \langle \nu'\| z \| \nu \rangle ,
\eeq
in which $(\nu \nu ,\, 2\Delta_1, 2\Delta_k | \nu'\nu')$ 
is a U$(N)$ vector-coupling coefficients, and
$\langle \nu'\| z \| \nu \rangle$ is U$(N)$-reduced matrix element.
Thus, the matrix elements of any $z_{kl}$ are obtained by the further use of eqn.\ (\ref{eq:WEthm})
and because $\partial_{kl}$ is the Hermitian adjoint of $z_{kl}$ in a unitary representation, its matrix elements  are also obtained from the Hermiticity relationship (\ref{eq:Hermadjs}).  

\subsection{Heisenberg algebras related to type III Capelli algebra}
\label{sect:Capell3rep}

The Heisenberg algebra that  emerges in the contraction of an 
$\mathfrak{so}(2N,\mathfrak{C})$  Lie algebra contains a  set of antisymmetric variables and derivative operators
\beq   z_{ij}= -z_{ji}, \quad  \partial_{ij}= -\partial_{ji} =  \partial/\partial z_{ij} ,
 \quad i,j = 1, \dots, N.
\eeq
that satisfy the commutation relations
\beq [\partial_{ij}, z_{kl}] = \delta_{i,k}\delta_{j,l}- \delta_{i,l}\delta_{j,k}.
\eeq

The Bargmann space $\Bb^{(\frac12 N(N-1))}$ spanned by polynomials in the
$\frac12 N(N-1)$ independent variables $\{ z_{ij}, i < j\}$ carries the unique unitary irrep of this Heisenberg algebra.
It is also the module for a highly reducible representation of the 
$\mathfrak{gl}(N)\subset\mathfrak{so}(2N,\mathfrak{C}) $  subalgebra with elements
\beq E^{(N)}_{ij} = \sum_{s=1}^N z_{is} \partial_{js} .
\eeq
This representation is a sum of irreps $\bigoplus_\nu \{ \nu\} $ of highest weight 
$\nu \equiv \{ \nu_1, \nu_2, \dots ,\nu_N\}$  which,
because 
\beq [E^{(N)}_{ii}, z_{kl}]= (\delta_{i,k} +\delta_{i,l})z_{kl} \label{[eq:E_ij,z_kl]} \eeq
implies that $z_{kl}$ is of weight $\Delta_k+\Delta_l$.

It will now be shown that the Bargmann space $\Bb^{(\frac12 N(N-1))}$ of polynomials in the antisymmetric $\{ z_{kl}\}$ variables is a sum of irreducible subspaces of highest weights $\nu = \{\nu_1,\nu_2, \dots , \nu_N\}$ with
\beqa 
&&\nu_1= \nu_2 \geq  \nu_3= \nu_4 \geq\dots \geq  
\nu_{N-1}= \nu_N  \geq 0  , \quad \text{if $N$ is even,} \\
&&\nu_1= \nu_2 \geq  \nu_3= \nu_4 \geq\dots \geq  
\nu_{N-2}= \nu_{N-1}  \geq \nu_N=0  , \quad \text{if $N$ is odd.}
\eeqa

\subsubsection{Highest-weight states for  irreps of ${\rm U}(N)$ on  subspaces of 
$\Bb^{(\frac12 N(N-1))}$}

As in sections \ref{sect:Capell1rep} and \ref{sect:Capell2rep},
we  define  a sequence of determinants 
\beq x_n := \det(z_{ij}) , \quad \nabla_n := \det(\partial_{ij}) ,
\quad i,j = 1, \dots , n ,\label{eq:xn&nablan3} \eeq
for each positive integer $n \leq N$.
However,  $x_1=z_{11}$ is now  identically zero and  the polynomial of highest 
$\mathfrak{u}(N)$ weight in the space of linear functions of the  $\{ z_{ij}\}$ variables is  $\phi_1(z) =z_{12}$.
Moreover, $\phi_1(z)$ is of weight $\{ 1^2\} \equiv \{1,1,0, \dots, 0\}$ and is a square root of the  determinant $x_2 = z_{12}^2$. 
In fact, it is known \cite{Parameswaran54,Lederman93} that every determinant of an $n\times n$ antisymmetric matrix  is identically zero when $n$ is odd.
However, when $n=2m$ is even, 
\beq x_{2m} = \phi_m^2 \eeq
is the square of the Pfaffian \cite{Weyl46}
\beq \phi_m(z) = \frac{1}{2^m m!} 
\sum_{\sigma\in {\rm S}_{2m}} {\rm sgn}(\sigma) z_{\sigma_1 \sigma_2}
 z_{\sigma_3 \sigma_4} \dots z_{\sigma_{2m-1} \sigma_{2m}}. \label{eq:so2N_h.wt}
 \eeq
It is also determined from eqn.\ (\ref{[eq:E_ij,z_kl]})
that $\phi_m$ is a polynomial of highest $\mathfrak{u}(N)$ weight
$\{ 1^{2m}, 0, \dots\}$.
Thus, the Pfaffians $\{ \phi_m\}$ are a complete set of generators of 
$\mathfrak{gl}(N)$ and $\mathfrak{u}(N)$ highest-weight polynomials given by the products
  \beq \Phi^{(2m)}_\nu =  \phi_1^{\nu_2-\nu_4}  \phi_2^{\nu_4-\nu_6} \dots
 \phi_{m-1}^{\nu_{2m-2} - \nu_{2m}} \phi_m^{\nu_{2m}} , \quad
 m= 1, \dots, \lfloor {N/2}\rfloor . \label{eq:Phi^2m_nu}
 \eeq
 The Hermitian adjoints of the Pfaffians are  the derivative operators
\beq \square_m = \frac{1}{2^m m!} 
\sum_{\sigma\in {\rm S}_m} {\rm sgn}(\sigma) \partial_{\sigma_1 \sigma_2}
 \partial_{\sigma_3 \sigma_4} \dots \partial_{\sigma_{2m-1} \sigma_{2m}} .
 \eeq

\subsubsection{Normalization factors for the  ${\rm U}(N)$ highest-weight states}

The type III Capelli identity \cite{HoweU91} states that the product of the determinants 
$x_n$ and $ \nabla_n$ satisfies the expression
\beq x_n \nabla_n = \det [E^{(n)}_{ij} +(n-1-i)\delta_{i,j}] , \label{eq:Capelli3a}
\eeq
where
\beq E^{(n)}_{ij} = \sum_{s=1}^n z_{is} \partial_{js} , \quad i,j = 1, \dots , n,
\label{eq:Eij(n)2}
\eeq
is an element of a $\mathfrak{u}(n)$ Lie algebra
and it is understood that the determinant on the right side of (\ref{eq:Capelli3a}) is defined with products of elements from different columns that preserve their  left to right order.
An  equivalent expression is  given by
\beq \nabla_n  x_n =
\det [E^{(n)}_{ij} +(n+1-i)\delta_{i,j}] . \label{eq:Capelli3b}
\eeq
However, because the highest-weight polynomials are now expressed in terms of Pfaffians rather than determinants,
this identity does not provides a direct determination of the squared norms
\beq  \mathcal{N}^{(2m)}_\nu = 
\langle \Phi^{(2m)}_\nu |\Phi^{(2m)}_\nu\rangle ,
\label{eq:normdefn3}  \eeq

There is no known expansion of the product $\square_m \phi_m$ similar to that given by the Capelli identity (\ref{eq:Capelli3b}) for $\nabla_n  x_n$.
However, it is known  \cite{LeBlancR87} that the highest-weight polynomials 
$\{ \Phi^{(2m)}_\nu \}$ are eigenfunctions of $\square_m \phi_m$,
\beq \square_m\phi_m\Phi^{(2m)}_\nu  =X_\nu \Phi^{(2m)}_\nu .
\label{eq:Xnu1}
\eeq
 with eigenvalues   (determined by complicated algebraic manipulations)
\beq X_\nu = (\nu_1+2m-1)(\nu_3+2m-3) \dots (\nu_{2m-1}+1)
= \prod_{i=1}^m \frac{(\nu_{2i}+2m+1-2i)!}{(\nu_{21} + 2m -2i)!}
 . \label{eq:Xnu2}
\eeq
As we now show, the eigenvalues $\{ X_\nu\}$ given by (\ref{eq:Xnu2})
are easily derived from the Capelli identity (\ref{eq:Capelli3b})  for which
\beq \nabla_{2m} x_{2m}\Phi^{(2m)}_\nu =  \square_m^2 \phi_m^2 \Phi^{(2m)}_\nu =
\prod_{i=1}^{2m} (\nu_i +2m + 1 - i)\Phi^{(2m)}_\nu .\eeq

Because $\phi_m$ has U$(N)$ weight $\{1^{2m}\}$ for $2m \leq N$,  
eqn.\ (\ref{eq:Xnu1}) implies that
\beq \square_m\phi^2_m\Phi^{(2n)}_\nu  =X_\mu\phi_m \Phi^{(2n)}_\nu,
\quad \text{for } m \geq n,
\label{eq:Xnu+Delta}
\eeq
with 
\beq \mu_i = \nu_i + 1 , \quad i=1, \dots, 2m.       \eeq
It then follows that
\beq \square^2_m\phi^2_m\Phi^{(2n)}_\nu 
=X_\mu \square_m\phi_m\Phi^{(2n)}_\nu
= X_\mu X_\nu \Phi^{(2n)}_\nu, \quad \text{for } m \geq n,
\eeq
and, because $\nu_{2i} = \nu_{2i-1}$, the required identity
\beq X_\mu X_\nu = \prod_{i=1}^{2m} (\nu_i +2m + 1 - i) \eeq
is obtained with $X_\nu$ given by eqn.\ (\ref{eq:Xnu2}).\\

\noindent {\bf Claim III:}
 The squared norm of  the highest-weight polynomial  $\Phi^{(2m)}_\nu $ 
 for a $\mathfrak{u}(N)$ irrep on the Bargmann space $\Bb^{(\frac12 N(N-1))}$ is
 given by
 \beq {\cal N}^{(2m)}_\nu = \langle  \Phi^{(2m)}_\nu | \Phi^{(2m)}_\nu\rangle =  
\frac{ \prod_{i=1}^{m} (\nu_{2i}+2m-2i)!}
{\prod_{i=1}^{m-1}\prod_{j=i+1}^m (\nu_{2i}-\nu_{2j}+2j-2i) (\nu_{2i}-\nu_{2j}+2j-2i-1)} .
 \label{claim:III}
 \eeq

\noindent {\bf Proof:}
Observe that the above extension of the expression for
$\square_m\phi_m\Phi^{(n)}_\nu$ to 
$\square^2_m\phi^2_m\Phi^{(n)}_\nu$ for $m\geq n$   further extends to  
\beq \square^p_m\phi^p_m\Phi^{(2n)}_\nu =
\prod_{i=1}^m \frac{(\nu_{2i} + 2m + p -2i)!}{(\nu_{2i} + 2m-2i)!} \Phi^{(2n)}_\nu , \quad
\text{for } m\geq n . \label{eq:Nmp}   \eeq
It follows that
\beq \langle \square_1^{p_1} \phi_1^{p_1}\rangle = p_1! .\eeq
Because eqn.\ (\ref{eq:Nmp}) implies that
\beq \square_2^{p_2} \phi_2^{p_2} \phi_1^{p_1} = 
\frac{(p_1+p_2+2)!p_2!}{(p_1+2)!} \phi_1^{p_1} ,
\eeq
it also follows that
\beq 
\langle \square_1^{p_1}\square_2^{p_2} \phi_2^{p_2} \phi_1^{p_1} \rangle = 
\frac{(p_1+p_2+2)!p_2!}{(p_1+2)!}\langle\square_1^{p_1} \phi_1^{p_1} \rangle
= \frac{(p_1+p_2+2)!p_2!}{(p_1+2)(p_1+1)}  ,
\eeq
and that, for $\Phi^{(4)}_\nu = \phi_2^{p_2} \phi_1^{p_1}$ with
$\nu_1=\nu_2= p_1+p_2$ and $\nu_3=\nu_4= p_2$,
\beq \langle \Phi^{(4)}_{\nu} |\Phi^{(4)}_{\nu}\rangle =
\frac{(\nu_2+2)! \nu_4!}{(\nu_2-\nu_4+2)(\nu_2-\nu_4+1)}
\eeq
as predicted by the claim.
Now,  if the claim is valid for  ${\cal N}^{(2m-2)}_\mu$ for some value of $m$, then
\beq 
 {\cal N}^{(2m)}_\nu =
 \langle \phi_m^{\nu_{2m}} \Phi^{(2m-2)}_\mu 
 | \phi_m^{\nu_{2m}} \Phi^{(2m-2)}_\mu \rangle 
 =\langle  \Phi^{(2m-2)}_\mu |
 \square_m^{\nu_{2m}} \phi_m^{\nu_{2m}} \Phi^{(2m-2)}_\mu \rangle  \label{eq:M(m-1)}
 \eeq
 with 
 \beq \nu_{2i} = \mu_{2i}+ \nu_{2m} .\eeq
Equations  (\ref{eq:Nmp}) and (\ref{eq:M(m-1)})  then give
\beqa 
 {\cal N}^{(2m)}_\nu&=&
\prod_{i=1}^m \frac{(\nu_{2i}+2m-2i)!}{(\nu_{2i}-\nu_{2m} + 2m-2i)!} \\
&&\quad \times
\frac{ \prod_{i=1}^{m-1} (\nu_{2i}-\nu_{2m}+2m-2i)!}
{\prod_{i=1}^{m-2}\prod_{j=i+1}^{m-1} (\nu_{2i}-\nu_{2j}+2j-2i) (\nu_{2i}-\nu_{2j}+2j-2i-1)} 
\\
&=& \frac{ \prod_{i=1}^{m} (\nu_{2i}+2m-2i)!}
{\prod_{i=1}^{m-1}\prod_{j=i+1}^{m} (\nu_{2i}-\nu_{2j}+2j-2i) (\nu_{2i}-\nu_{2j}+2j-2i-1)} ,
\eeqa
as claimed. 
Thus,  given that the claim is valid for $m\leq 2$, 
it is value for all $m$. 
 \hfill $\square$\\

\subsubsection{Matrix elements a type III Heisenberg algebra between
${\rm U}(N)$ highest-weight states}

  Having determined the  normalised highest-weight states 
\beq  |\nu\rangle := \frac{| \Phi^{(2m)}_\nu\rangle}
{\left[ \langle \Phi^{(2m)}_{\nu } | \Phi^{(2m)}_{\nu}\rangle \right]^{\frac12}} , 
\quad m =1, \dots , \lfloor N/2\rfloor
\eeq
for $\mathfrak{u}(N)$ irreps on $\Bb^{(\frac12 N(N-1))}$, 
all matrix elements of the type III Heisenberg algebra unitary irrep  
are determined, in terms of $\mathfrak{u}(N)$
vector-coupling coefficients, from the non-zero matrix elements
\beq \langle \nu' | z_{2k-1,2k} |\nu\rangle 
= \langle \nu  | \partial_{2k-1,2k} |\nu '\rangle^*,
 \label{eq:MEdeltakk3}
\eeq
where $z_{2k-1,2k}$ is the element of the Heisenberg algebra with weight
\beq \Delta_{2k-1} + \Delta_{2k} = \nu' - \nu .\eeq

By use of eqn.\ (\ref{eq:Phi^2m_nu}), the highest-weight wave functions are conveniently expressed in the form
\beq \Phi^{(2m)}_\nu = \phi^{p_1}_1 \phi^{p_2}_2 \dots \phi^{p_m}_m \eeq
with
$\nu_{2i-1} = \nu_{2i} =  \sum_{j=i}^m p_j$ . 
Then, with $\nu'_{2i-1} = \nu'_{2i} = \nu_{2i}+\delta_{i,k}$, we have
 \beq
\Phi^{(2m)}_{\nu'} = 
\phi^{p_1}_1  \dots \phi^{p_{k-2}}_{k-2} \phi^{p_{k-1}-1}_{k-1}
\phi^{p_k+1}_k  \phi^{p_{k+1}}_{k+1}   \dots  \phi^{p_m}_m  .
\eeq

The ratio 
$R= \langle \Phi^{(2m)}_{\nu} | 
\partial_{2k-1,2k} \Phi^{(2m)}_{\nu'}\rangle /
 \langle \Phi^{(2m)}_{\nu'} | \Phi^{(2m)}_{\nu'}\rangle$
has expansion
\beq R=\frac{
\langle \square_1^{p_1} \dots \square_k^{p_k} \partial_{2k-1, 2k}   W 
\phi_k^{p_k+1} \phi_{k-1}^{p_{k-1}-1} \phi_{k-2}^{p_{k-2}} \dots \phi_1^{p_1} \rangle}
{\langle \square_1^{p_1} \dots \square_{k-2}^{p_{k-2}}  \square_{k-1}^{p_{k-1}-1}
\square_{k}^{p_{k}+1}
   W 
\phi_k^{p_k+1}\phi_{k-1}^{p_{k-1}-1}\phi_{k-2}^{p_{k-2}} \dots \phi_1^{p_1} \rangle},
\eeq
where 
\beq   W = \square_{k+1}^{p_{k+1}} \dots \square_m^{p_m} 
\phi_m^{p_m} \dots \phi_{k+1}^{p_{k+1}} .
\eeq
Now, because the monomial to the right of the operator $  W$ is an eigenfunction of $  W$ and is the same in both the numerator and the denominator of this expression, $  W$ can be factored out of this ratio to give
\beqa 
R&=&
\frac{
\langle \square_1^{p_1} \dots \square_k^{p_k} 
[\partial_{2k-1,2k} , \phi_k^{p_k+1}] \phi_{k-1}^{p_{k-1}-1}\phi_{k-2}^{p_{k-2}} \dots \phi_1^{p_1} \rangle}
{\langle \square_1^{p_1} \dots \square_{k-2}^{p_{k-2}}  \square_{k-1}^{p_{k-1}-1}
\square_{k}^{p_{k}+1} 
\phi_k^{p_k+1}\phi_{k-1}^{p_{k-1}-1} \phi_{k-2}^{p_{k-2}} \dots \phi_1^{p_1} \rangle}
\\
&=& (p_k+1) \frac{
\langle \square_1^{p_1} \dots \square_k^{p_k} 
 \phi_{k}^{p_{k}}  \dots \phi_1^{p_1} \rangle}
{\langle \square_1^{p_1} \dots \square_{k-2}^{p_{k-2}} \square_{k-1}^{p_{k-1}-1}
\square_{k}^{p_{k}+1} 
\phi_k^{p_k+1}\phi_{k-1}^{p_{k-1}-1} \phi_{k-2}^{p_{k-2}} \dots \phi_1^{p_1} \rangle}
\\
&=& (p_k+1) \frac{\langle \Phi^{(k)}_\mu | \Phi^{(k)}_\mu\rangle}
{\langle \Phi^{(k)}_{\mu'} |  \Phi^{(k)}_{\mu'}\rangle},
\eeqa
where
\beq \Phi^{(k)}_\mu = 
 \phi_1^{p_1} \phi_2^{p_2} \dots  \phi_{k-1}^{p_{k-1}}\phi_k^{p_k}, \quad
\Phi^{(k)}_{\mu'} = 
 \phi_1^{p_1} \phi_2^{p_2} \dots  \phi_{k-1}^{p_{k-1}-1}\phi_k^{p_k+1}   ,
 \eeq
with  $\mu_{2i} = \sum_{j=i}^k p_j$ and 
$\mu'_{2i-1} = \mu'_{2i} = \mu_{2i}+\delta_{i,k}$.

Finallly, the desired matrix elements  
\beq \langle \nu'|  z_{2k-1,2k} |\nu\rangle = \frac
{\langle \Phi^{(m)}_{\nu'} | z_{2k-1,2k} \Phi^{(n)}_\nu\rangle}
{\left[ \langle \Phi^{(m)}_{\nu'} | \Phi^{(m)}_{\nu'}\rangle
\langle \Phi^{n}_{\nu } | \Phi^{n}_{\nu}\rangle \right]^{\frac12}} ,
\eeq
relative to normalized states, are given by
\beq \langle \nu'|  z_{2k-1,2k} |\nu\rangle = (\nu_{2k}-\nu_{2k+2}+1)
\frac{\langle \Phi^{(k)}_\mu |\Phi^{(k)}_\mu\rangle }
{\langle \Phi^{(k)}_{\mu'} |\Phi^{(k)}_{\mu'}\rangle}
\left[ \frac{\langle \Phi^{m}_{\nu' } | \Phi^{m}_{\nu'}\rangle}
{\langle \Phi^{m}_{\nu } | \Phi^{m}_{\nu}\rangle}\right]^{\frac12} \label{eq:UNMEs3}
\eeq
and easily evaluated with the expression for the norms given by Claim III.

\subsubsection{Unitary representation of the type III Heisenberg algebra
in a ${\rm U}(N)$-coupled  basis}

An element $z_{kl}$, with $k\not= l$, of the type III Heisenberg algebra  
is a component of  weight $\Delta_k +\Delta_l$ of a U$(N)$ tensor of  highest weight $\Delta_1+\Delta_2 =(1,1, 0, \dots )$.
Equation (\ref{eq:UNMEs3}) gives a non-vanishing matrix element 
$\langle \nu' | z_{2k-1,2k} |\nu\rangle$ with $k$ such that 
$\Delta_{2k-1}+ \Delta_{2k} = \nu' -\nu$.
Thus, according to the Wigner-Eckart theorem, 
\beq \langle \nu' | z_{2k-1,2k} |\nu\rangle = 
(\nu \nu ,\, \Delta_1+\Delta_2, \nu'-\nu | \nu'\nu') \langle \nu'\| z \| \nu \rangle ,
\eeq
where $(\nu \nu ,\, \Delta_1+\Delta_2, \nu'-\nu | \nu'\nu')$ 
is a U$(N)$ vector-coupling coefficients, and
$\langle \nu'\| z \| \nu \rangle$ is a U$(N)$-reduced matrix element.
Thus, the matrix elements of any $z_{kl}$ are obtained by the further use of eqn.\ (\ref{eq:WEthm})
and because $\partial_{kl}$ is the Hermitian adjoint of $z_{kl}$ in a unitary representation, its matrix elements  are also obtained from the Hermiticity relationship (\ref{eq:Hermadjs}).

\section{Summary and application of the results} \label{sect:conc}

The Capelli identities have been used to {construct the representations} of three classes of Heisenberg algebras: \\

(A)  Heisenberg algebras with  elements 
$\{ z_{i\alpha}, \partial _{i\alpha}; i= 1,\dots, p, \alpha = 1, \dots, q\}$ and commutation relations
\beq [z _{i\alpha}, z_{j\beta}] =[\partial _{i\alpha}, \partial_{j\beta}] =0, \quad
[\partial _{i\alpha}, z_{j\beta}] = \delta_{i,j} \delta_{\alpha,\beta}, \eeq
in bases that reduce the representations of the Lie algebras $\mathfrak{u}{(p)}$ and
$\mathfrak{u}{(q)}$ with elements given, respectively, by
\beq E^{(p)}_{ij} = \sum_{\alpha = 1}^q z_{i\alpha} \partial_{j\alpha},
\quad  E^{(q)}_{\alpha\beta} = -\sum_{i = 1}^p z_{i\beta} \partial_{i\alpha} ;
\eeq
(B) symmetric Heisenberg algebras with elements 
$\{ z_{ij}= z_{ji}, \partial_{ij}=\partial_{ji} ; i,j= 1,\dots, N\}$ and commutation relations
\beq [z _{ij}, z_{kl}] =[\partial_{ij}, \partial_{kl}] =0, \quad
[\partial_{ij}, z_{kl}] = \delta_{i,k} \delta_{j,l} +\delta_{i,l} \delta_{j,k}, \eeq
in bases that reduce the representation of $\mathfrak{u}{(N)}$  with elements 
\beq E^{(N)}_{ij} = \sum_{s = 1}^N z_{is} \partial_{js} ;
\eeq
(C) antisymmetric Heisenberg algebras with elements 
$\{ z_{ij}= -z_{ji}, \partial_{ij}=-\partial_{ji}; i,j= 1,\dots, N\}$ and commutation relations
\beq [z _{ij}, z_{kl}] =[\partial _{ij}, \partial_{kl}] =0, \quad
[\partial _{ij}, z_{kl}] = \delta_{i,k} \delta_{j,l} -\delta_{i,l} \delta_{j,k}, \eeq
in bases that likewise reduce the representation of $\mathfrak{u}{(N)}$  with elements 
\beq E^{(N)}_{ij} = \sum_{s = 1}^N z_{is} \partial_{js} .
\eeq

Note that by ``constructing a representation'' we mean much more than simply defining a module for the representation as explained in the introduction. 
Constructions are also given for  elementary irreps on a Bargmann space 
$\Bb$  of the semi-direct sums of  these Heisenberg algebras and the associated unitary Lie algebras spanned, for example, by  elements
$\{ E^{(N)}_{ij}, z_{ij} , \partial_{ij} ,I, i,j = 1, \dots , N\}$, where $I$ is the identity element of the Heisenberg algebra.
Moreover, they define asymptotic limits of the holomorphic representations of the Lie algebras from which the semi-direct sums are obtained by contraction.
For example, for suitably chosen values of $\{ k^{(\kappa)}\}$, 
the contraction of $\mathfrak{sp}(N,\Rb)$ given by  
eqn.\ (\ref{eq: contractedSp}) leads to an asymptotic limit
\beq  C_{ij} \to\Cb^{(\kappa)}_{ij} +\sum_s   z_{is} \partial_{js} , \quad
A_{ij} \to k^{(\kappa)} z_{ij}, \quad  B_{ij} \to k^{(\kappa)*} \partial_{ij} .
\eeq
of an  $\mathfrak{sp}(N,\Rb)$ irrep  for which the lowest-weight state 
belongs to an irrep $\{ \kappa\}$ of the 
$\mathfrak{u}(N) \subset\mathfrak{sp}(N,\Rb)$ subalgebra. 
This asymptotic limit is defined on a tensor product space
$\mathcal{F}^{(\kappa)} := \Bb^{(\frac12 N(N+1))}\otimes \Hb_0^{\kappa}$ of holomorphic vector-valued functions, 
where $\Hb^{(\kappa)}_0$ is the Hilbert space for the irrep 
$\{ \kappa\}$ of  $\mathfrak{u}(N)$, on which the operators 
$\Cb^{(\kappa)}_{ij}$ act.

Such asymptotic limits  correspond to macroscopic limits of microscopic models in physics. 
They also lead to a  construction of the holomorphic representations of the Lie algebras from which the contracted Lie algebras derive.
For example, vector coherent state (VCS) theory \cite{Rowe84,RoweR91,Rowe12}.
shows that a  holomorphic vector-valued representation of 
$\mathfrak{sp} (N,\Rb)$ 
\beq
\Gamma^{(\kappa)}(C_{ij})= \Cb^{(\kappa)}_{ij}  
+ \sum_{k=1}^q z_{ik} \partial z_{jk} , 
 \quad
\Gamma^{(\kappa)}( A_{ij} ) = [ \Lambda, z_{ij}] ,\quad 
\Gamma^{(\kappa)}( B_{ij}) =  \partial_{ij},  \label{eq:spNR.holo1}
\eeq
is  obtained on the same Hilbert space  $\mathcal{F}^{(\kappa)}$ as the contracted irrep where $\Lambda$ is a U$(N)$-invariant Hermitian operator
that has known eigenvalues on the U$(N)$-coupled basis for 
$\mathcal{F}^\kappa$.

The holomorphic representations obtained from contractions in this way are precise, as opposed to approximations given by asymptotic limits that are often  appropriate in applications to large many-particle systems.
They are not unitary with respect to an orthonormal basis for a unitary representation of the contracted algebras.  
Neither, in general, are they irreducible.
However, VCS theory includes a K-matrix  procedure 
\cite{Rowe84,Rowe95,Rowe12} for orthogonalizing and 
renormalizing the bases to give explicit  unitary representations on an  
irreducible subspace of  $\mathcal{F}^\kappa$.
This procedure is simplified by a recent observation of 
an equivalent  VCS representation,  $ \Theta^{(\kappa)}$, 
of the $\mathfrak{sp}(N,\Rb)$ algebra  of the form
\beq
\Theta^{(\kappa)}(C_{ij})= \Cb^{(\kappa)}_{ij}   
+ \sum_{k=1}^q z_{ik} \partial z_{jk} , 
 \quad
\Theta^{(\kappa)}( A_{ij} ) =z_{ij} ,\quad 
\Theta^{(\kappa)}( B_{ij}) =  [\partial_{ij},\Lambda] , \label{eq:spNR.holo2}
\eeq
which, like $\Gamma^{(\kappa)}$, 
is defined on the Hilbert space 
$\mathcal{F}^\kappa$ for a unitary representation of the contracted Lie algebra.  
In fact, as will be shown explicitly in ref.\ \cite{Rowe15}, $\Theta^{(\kappa)}$ 
and $\Gamma^{(\kappa)}$ are
 bi-orthogonal duals of one another relative
 to the  inner product appropriate for the contracted  $\mathfrak{sp}(N,\Rb)$ algebra.
Thus, from the paired representations 
$\Theta^{(\kappa)}$ and $\Gamma^{(\kappa)}$, 
it is a straightforward procedure to construct orthonormal bases and irreducible unitary holomorphic representations of all the standard Lie algebras with holomorphic representations.

A  practical advantage of the K-matrix procedure of VCS theory is that it  makes use of algebraic methods to determine an orthonormal basis.
It thereby avoids use of the standard integral form of the inner product, 
based on a resolution of the identity, 
which converges only for discrete series representations \cite{Rowe15}.  
In contrast, the VCS construction of holomorphic representations is not restricted to discrete series representations.

\begin{acknowledgements}
Helpful suggestions from R.\ Howe, T.\ Umeda,  E.\ M.\ Baruch and
J. Repka are gratefully acknowledged.  The author is particularly appreciative of many constructive suggestions from the referee.

\end{acknowledgements}

\bibliographystyle{apsrev}
\bibliography{Jrw-book2}

\end{document}